\documentclass[aps,prb,twocolumn,groupedaddress,footinbib]{revtex4-2}

\usepackage{amsmath}
\usepackage[dvips]{color}
\usepackage[utf8]{inputenc}
\usepackage[american]{babel}
\usepackage{color}
\usepackage{graphicx}
\usepackage{epstopdf}
\usepackage{epsfig}
\usepackage{float}
\usepackage[colorlinks,linkcolor=blue,citecolor=blue,urlcolor=blue]{hyperref}
\usepackage{comment}
\usepackage{amssymb}
\usepackage{multirow}
\usepackage[table]{xcolor}
\usepackage[normalem]{ulem}

\newcommand{\angstrom}{\mbox{\normalfont\AA}}

\newcommand{\IM}{\mbox {Im}}

\allowdisplaybreaks

\begin{document}

\title{Importance of coupling strength in shaping electron energy loss and phonon spectra of phonon-plasmon systems}

\author{J. Krsnik}
\email{jkrsnik@ifs.hr}
\affiliation{Department for Research of Materials under Extreme Conditions, Institute of Physics, HR-10000 Zagreb, Croatia}

\author{O. S. Bari\v si\' c}
\affiliation{Department for Research of Materials under Extreme Conditions, Institute of Physics, HR-10000 Zagreb, Croatia}

\begin{abstract}
	
{A systematic analysis of phonon-plasmon coupled excitations in three-dimensional (3D) polar systems is provided through the prism of both raw and integrated electron energy loss spectroscopy (EELS) and phonon spectra in the whole relevant parametric space, spanned by the adiabaticity parameter and the electron-phonon interaction (EPI) strength. We show that the EPI strength plays a prominent role in distributing spectral weights among excitations, providing an experimentally convenient way to estimate it from integrated spectra. By projecting the excitations onto the phonon degree of freedom, we also report for strong couplings large phonon production contributions, which are of very different origins depending on the adiabaticity parameter. In parallel to this thorough spectral weights analysis, excitations' dispersion evolutions, dampings, and various limiting behaviors are qualitatively and quantitatively correctly accounted for in the whole parametric space.}
	
\end{abstract}

\maketitle 

\section{Introduction}
The electron-phonon interaction (EPI) accounts for a great variety of enthralling physical phenomena observed in condensed matter systems. In semiconductors with low electron densities, it can lead to a polaron formation where itinerant charges move along the crystal lattice dressed by a cloud of virtual phonons \cite{Landau,emin,barisic2008}. It has also been recognized that in a weakly doped semiconductor the electron and the phonon scattering on a polaronic impurity may greatly affect transport properties of a material \cite{mettan,krsnik2020}. Peculiar effects of the EPI are also prominent in systems with high electron densities. In  metallic systems, for instance, the EPI may result in a transition to a conventional BCS superconducting \cite{schrieffer} or a charge density wave state  \cite{zhu2015,pouget,hohenadler2018}. Influences of the EPI are noticeable in spectral features of heavily doped semiconductors as well. While the photoemission spectra feature phonon sidebands below the quasiparticle band \cite{moser,cancellieri2016,verdi2017,caruso2018,krsnik}, both the Raman and the infrared spectroscopy measurements on doped polar semiconductors contain fingerprints of a phonon-plasmon coupling \cite{mooradian,abstreiter,bell1968,murase1974,romanek1981,fukasawa1994,olson1969,kukharskii1973,chandrasekhar1980,gervais1993,bi2006,radovic2015}.

The coupling of a longitudinal optical (LO) phonon with a longitudinal plasma oscillation in three-dimensional (3D) polar semiconductors was addressed as early as the 1960s in seminal works by Yokota \cite{yokota}, Varga \cite{varga}, Singwi \textit{et al}. \cite{singwi1966}, and Cochran \textit{et al}. \cite{cochran1966}.
By varying the electron density, the frequency of long-wavelength plasma oscillation can be tuned, allowing it to be in a resonance with the phonon frequency, which results in a level repulsion of two boson modes. This phenomenon was oftentimes observed experimentally in the Raman \cite{mooradian,abstreiter,bell1968,murase1974,romanek1981,fukasawa1994} and infrared spectroscopy \cite{olson1969,kukharskii1973,chandrasekhar1980} measurements of a highly doped GaAs, as well as of some transition metal oxides \cite{gervais1993,bi2006,radovic2015}, giving an excellent agreement with theoretical predictions \cite{inaoka1991}. In the past few years, the level repulsion of modes was captured in ultrafast transient reflectivity measurements of III-IV semiconductors \cite{ishioka2011,basak2015,ishioka2015,hu2018} as well, leading to the renewed interest in studying the phonon-plasmon coupled excitations in 3D systems.

Apart from the examination of level repulsion in the long-wavelength limit, several studies analyzed the dispersions of coupled modes outside the electron-hole continuum, examined by considering zeros of the longitudinal dielectric function or extracted from the simulated electron energy loss spectroscopy (EELS) spectra \cite{varga,cochran1966,inaoka1991,lemmens1974,lemmens1975}. The level of influence of the continuum on coupled modes remained thus unclear; there were attempts to resolve this by considering scattering of electrons on collective excitations \cite{kim1978,sato1987}, albeit definite answer about excitations' damping was still not provided.

Another important question of the excitations' character was addressed in \cite{inaoka1991} and \cite{sato1987} by considering phonon and plasmon strengths in the total dielectric function and EELS spectra, respectively. These results were further supplemented by considering phonon strengths of coupled modes in the long-wavelength limit \cite{varga,singwi1966,kim1978}. However, the full phonon spectrum in the presence of the phonon-plasmon coupling was seldom analysed. We may mention a work by Yi \textit{et al}. \cite{yi2015}, which, however, overlooks the spectral weight of collective excitations outside the continuum and lacks the very interesting resonant regime. Actually, in all of the works, to the best of our knowledge, only the adiabaticity parameter, i.e., the electron density, was recognized as a quantity which separates quantitatively different coupling regimes, while the influence of the EPI strength was completely overlooked.

In this paper, we provide a systematic and thorough analysis of phonon-plasmon coupled excitations in the whole two-dimensional parametric space, spanned by both the adiabaticity parameter and the EPI strength. In particular, we distinguish six very different regimes, which come as a product of three adiabaticity regimes: (a) antiadiabatic, (b) resonant, and (c) adiabatic, and two EPI strength cases: (A) weak and (B) strong. Characteristics of all the regimes are studied by means of fully momentum and energy resolved raw and integrated EELS and phonon (corresponding to the LO phonon) spectra, providing for the first time a side by side comparison of coupled excitations' projections onto the electron and the phonon degree of freedom. To cover all the parametric space, we refer to two polar semiconducting materials that are of particular interest, the bulk GaAs, which is on the weak EPI side of the phase diagram, and TiO$_2$ with the considerably stronger EPI.

As anticipated, the damping of excitations by the continuum is enhanced as the EPI strength increases. However, here we report that in some regimes one collective excitation may stay coherent throughout the whole continuum, even for strong coupling strengths. This is manifested as a weak spread of the excitation's weight among the continuum in raw spectra, which as well guide us to correctly account for excitations' dispersions across the whole continuum.

Differences between the weak and the strong coupling cases become even more apparent when integrated spectra are considered. In particular, the strong EPI favors the total EELS spectral weight confined only within the single excitation, while for the weaker coupling both coupled excitations may share appreciable EELS spectral weight. This renders the integrated EELS spectra a very convenient tool for the experimental analysis of the phonon-plasmon coupling, particularly the EPI strength, even with the current energy resolution limitations in experiments. Namely, at the present time the energy resolution $\Delta E \sim 30 - 100$ meV is not high enough to fully spectrally resolve characteristic frequencies of coupled modes. Accordingly, the raw EELS measurements cannot be efficiently exploited to study the phonon-plasmon coupled excitations, like for example highly energetic plasmons in metals \cite{powell1959al,powell1959mg,nagao2007}, and, recently, in heavily doped semiconductors \cite{granerod2018,yang2020}. While new techniques that overcome these technical difficulties are appearing \cite{krivanek2014}, our analysis based on integrated spectra may circumvent the problem of the limited experimental resolutions to a great extent.

Besides the standard Raman and the infrared spectroscopy for momenta close to the center of the Brillouin zone, an alternative with a sufficient energy resolution to capture dispersions of phonon-plasmon coupled modes \cite{Smolyaninova}  is experiments based on the inelastic neutron scattering \cite{brockhouse1955,strauch1990}, highlighting the importance of the projection of coupled excitations on the phonon degree of freedom. By conducting the in-depth analysis of phonon spectral features, we show that the strong-coupling case is accompanied by large phonon production contributions, which makes us especially emphasize the importance of a distinction between phonon softening effects and effects caused by a virtual cloud of phonons attached to charge fluctuations.

\section{General}

We analyze a single band model that describes a semiconductor with a bottom/top of conduction/valence band doped, such that a dispersion of electrons/holes may be assumed quadratic $\varepsilon_{\mathbf{k}} = \frac{\hbar^2k^2}{2m^*}$, characterized by an effective mass $m^*$. In addition to the Coulomb interaction between itinerant charges, we investigate effects of the electron/hole interaction with lattice phonons, assuming that the latter is dominated by the polar coupling to a dispersionless LO phonon  branch with the frequency $\omega_{LO}$. 

Our model Hamiltonian for bulk 3D materials is given by \cite{mahan}

\begin{equation}
\begin{split}
	\hat{H} &= \sum_{\mathbf{k}}^{}\varepsilon_{\mathbf{k}}c^\dagger_{\mathbf{k}}c_{\mathbf{k}} + \hbar\omega_{LO}\sum_{\mathbf{k}}^{}a^\dagger_{\mathbf{k}}a_{\mathbf{k}} + \frac{1}{2}\sum_{\mathbf{q}}^{}v_\mathbf{q}^{\infty}\hat{\rho}_\mathbf{q}\hat{\rho}_{-\mathbf{q}}\\
	& + \sum_{\mathbf{q}}M_\mathbf{q}\hat{\rho}_\mathbf{q}\left[  a^\dagger_{\mathbf{q}}+a_{\mathbf{-q}}\right]\;,\label{Ham}
\end{split}	
\end{equation}

\noindent where $c^\dagger_{\mathbf{k}}$ and $a^\dagger_{\mathbf{k}}$ are the creation operators of the electron/hole and the phonon with the wave vector $\mathbf{k}$, respectively, and $\hat{\rho}_\mathbf{q} = \sum_{\mathbf{k}}c^\dagger_{\mathbf{k}+\mathbf{q}}c_{\mathbf{k}}$  is the charge density operator.
The screening from high-energy excitations across band gaps is taken into account through the high-frequency dielectric constant $\varepsilon_\infty$,  characterizing the interaction between the electrons in Eq.~\eqref{Ham}, $v_\mathbf{q}^{\infty} = v_\mathbf{q}/\varepsilon_\infty$, with $v_\mathbf{q} =e^2/ \epsilon_0 q^{2}V$ the Coulomb potential, where $\epsilon_0$ is the vacuum permittivity. Due to this screening, the interband excitations renormalize the plasmon frequency, $\Omega_{PL}^\infty=\Omega_{PL}/\sqrt\varepsilon_\infty$, where $\Omega_{PL}=\sqrt{ne^2/\varepsilon_0m^*}$ would be the plasmon frequency in the absence of other bands, and $n$ is the density of itinerant charge carriers.

As far as the EPI  is concerned, corresponding to the last term in  Eq.~\eqref{Ham},  we assume a polar coupling described by the Fr\"ohlich model \cite{devreese2010}

\begin{equation} \label{eq:polarmatrixelement}
	M_\mathbf{q} = -i\sqrt{v_\mathbf{q}^{\infty}} \sqrt{\frac{\hbar \omega_{LO}}{2}} \sqrt{1 - \frac{\varepsilon_\infty}{\varepsilon_0}}\;.
\end{equation}

\noindent Here, $\varepsilon_0$ is the static dielectric constant of a polar crystal, measured well below the phonon frequency $\omega_{LO}$ (not to be confused with $\epsilon_0$). In the spirit of polaron theories for an electron doped in an empty band, we introduce a dimensionless electron-phonon coupling constant \cite{devreese2010}

\begin{equation}
	\alpha = \frac{e^2}{4\pi\epsilon_0\hbar}\sqrt{\frac{m^*}{2\hbar\omega_{LO}}}\left( \frac{1}{\varepsilon_\infty}-\frac{1}{\varepsilon_0}\right)\;,
\end{equation}

\noindent characterizing the strength of the EPI across the phase diagram. In particular, the values $\alpha\ll 1$ and $\alpha\approx 1$ correspond to the weak and the strong EPI case, respectively, as found in standard semiconducting materials. In the case of the polaron problem, $\alpha$ defines the leading contribution to the polaron binding energy, $E_{pol}\approx\alpha\;\hbar\omega_{LO}$. It should be emphasized that all the bare model parameters in Eq.~\eqref{Ham} may be determined either from experiments or by performing $ab\;initio$ calculations for the undoped polar semiconductor of interest.

\subsection{EELS spectrum}

The inelastic scattering cross section of electrons measured in an EELS experiment is related via the fluctuation-dissipation theorem to the system's charge density-density correlation function \cite{soliom}, and hence an EELS spectrum is directly proportional to the imaginary part of the inverse of the system's total dielectric function

\begin{equation}\label{eq:EELSspectrum}
\begin{split}
	S(\mathbf{q},\omega)&\propto -\pi^{-1}\text{Im}\left[ \varepsilon^{-1} (\mathbf{q},\omega)\right] \\&= \pi^{-1}\frac{\text{Im}\; \varepsilon(\mathbf{q},\omega)}{\left[ \text{Re} \;\varepsilon(\mathbf{q},\omega)\right]^2 +\left[ \text{Im}\;\varepsilon(\mathbf{q},\omega)\right]^2 }\;.
\end{split}
\end{equation} 

\noindent In order to simulate EELS spectra, in this paper we adopt the random phase approximation (RPA) kind of scheme for the total dielectric function of the system. Within this scheme, the electron and the phonon contributions to the dielectric function are additive, yielding \cite{mahan}

\begin{equation} \label{eq:totaldielectricfunction}
	\begin{split}
		\varepsilon(\mathbf{q},\omega) &= \varepsilon_\infty - v_{\mathbf{q}}\chi_0(\mathbf{q},\omega) + \varepsilon_\infty\frac{\omega_{pl}^2}{\omega_{TO}^2-\omega^2}\\
		& = \varepsilon_{\infty}\left[ \varepsilon_{RPA}(\mathbf{q},\omega)+\frac{\omega_{pl}^2}{\omega_{TO}^2-\omega^2}\right] \;.
	\end{split}
\end{equation}

\noindent The first term in the first row of Eq.~\eqref{eq:totaldielectricfunction} accounts for the high-energy interband  excitations, the second term for the intraband excitations, and the last term  for the phonon contribution. In the second row of Eq.~\eqref{eq:totaldielectricfunction}, we simply exploited the standard RPA form for the electron dielectric function

\begin{equation} \label{eq:electrondielectricfunction}
		\varepsilon_{RPA}(\mathbf{q},\omega) = 1 - v^\infty_{\mathbf{q}}\chi_0(\mathbf{q},\omega)\;.
\end{equation}
Here, as well as in Eq.~\eqref{eq:totaldielectricfunction}, $\chi_0(\mathbf{q},\omega)$ is the polarization bubble (the Lindhard function), contributed by electron-hole pair excitations

\begin{equation}
	\chi_0(\mathbf{q},\omega) = \frac{2}{V}\sum_{\mathbf{k}}\frac{f_{\mathbf{k}}-f_{\mathbf{k}+\mathbf{q}}}{\hbar\omega - \varepsilon_{\mathbf{k}+\mathbf{q}}+\varepsilon_{\mathbf{k}} + i\eta}\;.\label{Lindhard}
\end{equation}
The factor 2 accounts for the electron spin degeneracy, and $f_{\mathbf{k}}$ is the Fermi-Dirac distribution. In the zero-temperature limit analyzed in this work, $f_\mathbf{k}=1$ and $f_\mathbf{k}=0$ below and above the Fermi level, respectively. The frequency of the transversal optical (TO) phonon is given by the LO phonon frequency through the Lydanne-Sachs-Teller (LST) relation $\omega_{TO}^2=\omega_{LO}^2 \varepsilon_\infty/\varepsilon_0$, or through the ionic plasma frequency $\omega_{pl}^2 = \omega_{LO}^2-\omega_{TO}^2$ \cite{mahan}.

Outside the electron-hole continuum, $\text{Im}\;\varepsilon(\mathbf{q},\omega)$ is zero and $\text{Im}\left[ \varepsilon^{-1} (\mathbf{q},\omega)\right]$ has the  form of delta peaks at frequencies of system collective excitations $\omega_i$, with the corresponding spectral weights given by 

\begin{equation}
\left. s_i\left( \mathbf{q}\right)  = \left[ \frac{\partial \varepsilon(\mathbf{q},\omega)}{\partial \omega}\right]^{-1} \right|_{\omega=\omega_i}\;.
\end{equation}

\noindent  Once the continuum is reached, the EELS spectrum acquires an incoherent contribution from electron-hole pair excitations $s_{e-h}\left( \mathbf{q},\omega\right)$ as well. By integrating the spectrum over frequencies, one obtains the total spectral weight as a function of $\mathbf{q}$

\begin{eqnarray} \label{eq:eelsintegrated}
s_{tot}\left( \mathbf{q}\right)&=&\int_{0}^{+\infty}d\omega\left\lbrace -\pi^{-1}\text{Im}\left[ \varepsilon^{-1} (\mathbf{q},\omega)\right]\right\rbrace \nonumber\\
&=& \sum_{i}s_i\left( \mathbf{q}\right)+s_{e-h}\left( \mathbf{q}\right)\;.
\end{eqnarray} 

In the long-wavelength limit and in the absence of the EPI, the EELS spectrum is dominated by the plasmon excitation

\begin{equation} \label{eq:weightplasmoninfinity}
s_{tot}\left( q\to0\right)\approx s_{\Omega^{\infty}_{PL}}\left( q\to0\right)=\frac{\hbar \Omega^{\infty}_{PL}}{2\varepsilon_{\infty}}\;.
\end{equation}

\noindent With the introduction of finite EPI, the electrons scatter on phonons at the rate proportional to the strength of the EPI. As shown in Appendix \ref{weightsEELS}, the EELS spectral weight  at $\omega=\omega_{LO}$, characterizing this scattering, is given by

\begin{equation} \label{eq:weightLO}
s_{\omega_{LO}}\left( q\to0\right) = \frac{\hbar\omega_{LO}}{2}\left(\frac{1}{\varepsilon_\infty}-\frac{1}{\varepsilon_0} \right)\;. 
\end{equation}

\noindent However, for the itinerant charge concentrations when the plasmon and the phonon frequencies become comparable, the two kinds of excitations are strongly coupled, resulting with significantly renormalized energies of coupled excitations. They do not necessarily have predominantly a phonon or a plasmon character, but rather a hybridization of excitations is generally expected.  In such circumstances, in the long-wavelength limit the EELS spectrum shows two peaks at energies $E_-$ and $E_+$, hereafter referred to as to the lower frequency excitation (LFE) and higher frequency excitation (HFE), respectively. With $s_-$ and $s_+$ we denote the corresponding spectral weights of the two coupled excitations in EELS spectra.

\subsection{Phonon spectral function}

To discuss the mixing between the plasmon and the phonon nature of collective excitations, in parallel to the EELS spectrum it is very useful to consider an experimentally relevant dynamical quantity that involves a projection of system excitations on phonon degrees of freedom. In particular, here we analyze the phonon spectral function that may be investigated by neutron scattering experiments, defined by

\begin{equation}
	B(\mathbf{q},\omega) = -\pi^{-1} \text{Im}D(\mathbf{q},\omega)\;, 
\end{equation}

\noindent where $D(\mathbf{q},\omega)$ is the LO phonon propagator. The integrated spectral weight satisfies the sum rule given by

\begin{equation}
\int_{-\infty}^\infty B(\mathbf{q},\omega)d\omega=\frac{2M\omega_{LO}}{\hbar}\langle0|\hat x_\mathbf{q}\hat x_{-\mathbf{q}}|0\rangle\label{PHsumrule}
\end{equation}

\noindent where $\hat x_\mathbf{q}=(a_\mathbf{q}^\dagger+a_{-\mathbf{q}})\sqrt{\hbar/2M\omega_{LO}}$ is the lattice displacement operator, and $M$ the characteristic ion mass. According to Eq.~\eqref{PHsumrule}, the total spectral weight is defined by the lattice quantum fluctuations in the ground state of the interacting system  $|0\rangle$. In general, this spectral weight may be distributed among different excitations of the system.

The spectral function of the bare phonon propagator 

\begin{equation}
	D^{(0)}(\mathbf{q},\omega) = \frac{1}{\omega-\omega_{LO}+i\eta}-\frac{1}{\omega+\omega_{LO}-i\eta}\;,
\end{equation}
is characterized by the LO phonon frequency only,

\begin{equation} \label{eq:freespectral}
	B^{(0)}(\mathbf{q},\omega) = \left[\delta(\omega - \omega_{LO}) + \delta(\omega + \omega_{LO}) \right] \;.
\end{equation}

\noindent By accounting for interaction, the full phonon propagator may be expressed in terms of the phonon self-energy $\Pi(\mathbf{q},\omega)$ through the Dyson equation

\begin{equation} \label{eq:dyson}
	\left[ D(\mathbf{q},\omega)\right] ^{-1} = \left[ D^{(0)}(\mathbf{q},\omega)\right] ^{-1}-\Pi(\mathbf{q},\omega) \;.
\end{equation}
The phonon self-energy contains the both contributions, from the EPI and the electron-electron interaction. 

Since the phonon propagator is even in $\omega$, it is sufficient to analyze positive frequencies only. In the absence of interactions, $\Pi(\mathbf{q},\omega)=0$, the phonon spectral function Eq.~\eqref{eq:freespectral} satisfies the sum rule

\begin{equation} \label{eq:sumrule}
	\int_{0}^{+\infty} d\omega B^{(0)}(\mathbf{q},\omega) = 1\;,
\end{equation}

\noindent corresponding to the zero-point motion of the $\omega_{LO}$ phonon.

With the interactions included, the phonon spectral function develops new peaks corresponding to collective system excitations, given by the poles of Eq.~\eqref{eq:dyson}. This is accompanied by the appearance of an incoherent contribution from electron-hole pair excitations, yielding

\begin{equation}
	D(\mathbf{q},\omega>0) = \sum_{i}\frac{z_i(\mathbf{q})}{\omega-\omega_i(\mathbf{q}) + i\eta} + D_{e-h}(\mathbf{q},\omega>0)\;.
\end{equation}

\noindent Here, $z_i$ measures directly the projection of system excitations to the bare phonon, providing the information about their phonon character. The crucial observation to make is that in addition to the spectral weight redistribution among excitations and incoherent continuum, an additional spectral weight might appear,

\begin{equation} \label{eq:phononintegrated}
	\int_{0}^{+\infty} d\omega B(\mathbf{q},\omega) = \sum_{i}z_i(\mathbf{q}) + z_{e-h}(\mathbf{q})\geq 1\;,
\end{equation}

\noindent where $z_{e-h}(\mathbf{q},\omega>0)$ denotes the phonon spectral weight associated with the incoherent continuum. This additional spectral weight, which we call a phonon production, depends on the model parameters and itinerant charge concentration. It has been discussed in the context of polaron physics and the Huang scattering \cite{ranninger1992,barisic2006}, when it corresponds to the polaronic lattice deformation present in the ground state of EPI systems.

In analogy to the EELS spectra, due to the phonon-plasmon coupling, in the long-wavelength limit two excitations with finite energies, $E_\pm$, appear in the phonon spectral function. Each of these two excitations is characterized by their own spectral weight $z_\pm$. The cases with the phonon production, $z_+ + z_- > 1$, deserve a special attention since they may be caused by different physical mechanisms, ranging from the standard phonon softening to the presence of a permanent, yet dynamical lattice deformation which does not break the translational symmetry. In the context of phonon-plasmon coupled systems, to the best of our knowledge the phonon production has not been discussed previously, with the exception of the seminal work by Varga \cite{varga}.

\begin{figure}[h]	
	\centering
	\includegraphics[width=1.\columnwidth]{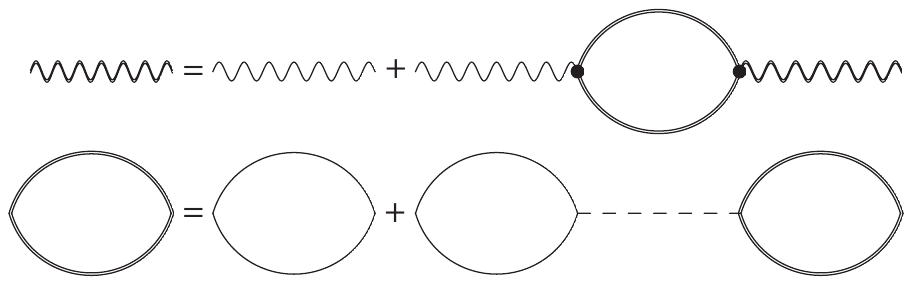}
	\renewcommand{\figurename}{FIG.}
	\caption{Diagrammatic representation of our RPA kind of scheme for the phonon propagator (double wavy line). The bare phonon propagator is represented with a single wavy line, while the phonon self-energy consists of the density-density correlation function (double bubble) ending with the EPI vertices represented by black dots. In the RPA, the density-density correlation function can be pictorially represented by the infinite chain of bubble diagrams coupled with the Coulomb interaction (dashed line), as shown in the second row.}
	\label{fig:RPAapprox}
\end{figure}

To treat on an equal footing the EELS and the phonon spectra, we adopt the RPA kind of scheme for calculations of the phonon propagator. Diagrammatically, this scheme is shown in Fig. \ref{fig:RPAapprox}. The first row in Fig.~\ref{fig:RPAapprox} represents the standard Dyson equation for the phonon propagator, with the phonon self-energy being represented by the bubble involving two double lines ending by the pair of electron-phonon vertices (dots). The bubble in the first row stands for the standard RPA approximation for the density-density correlation function in a system with the electron-electron interaction, shown in the second row in Fig.~\ref{fig:RPAapprox}, with the dashed line representing the instantaneous electron-electron interaction. It is easy to verify that the scheme in Fig.~\ref{fig:RPAapprox} corresponds to an infinite series of diagrams for the phonon propagator, where the series consists of the diagrams with all different number of polarization bubbles connected either with phonon or electron-electron interaction lines in all possible combinations. It is worth mentioning that the scheme in Fig. \ref{fig:RPAapprox} can be utilized to obtain the total dielectric function, Eq.~\eqref{eq:totaldielectricfunction}, simply by reinterpreting the double wavy line as the effective electron-electron interaction and interchanging the phonon (together with the EPI vertices) and the Coulomb interaction propagators.

By recalling that the electron dielectric function in the standard RPA takes the form given by Eq.~\eqref{eq:electrondielectricfunction}, the phonon self-energy corresponding to  Fig.~\ref{fig:RPAapprox} may be written as
 
\begin{equation} \label{eq:RPApolarization}
\begin{split}
	\Pi(\mathbf{q},\omega) = \frac{|M_\mathbf{q}|^2}{v^{\infty}_\mathbf{q}}\left[ \frac{1}{\varepsilon_{RPA}(\mathbf{q},\omega)}-1\right]\;. 
\end{split}	
\end{equation}
This expression is especially appealing, since it readily allows for the calculation of phonon self-energy beyond the RPA. Namely, replacing $\chi_0(\mathbf{q},\omega)$ with $\chi_{irr}(\mathbf{q},\omega)$ in $\varepsilon_{RPA}(\mathbf{q},\omega)$, where $\chi_{irr}(\mathbf{q},\omega)$ denotes the sum of irreducible polarization bubble diagrams, allows for the inclusion of self-energy and vertex corrections due to both the electron-electron and the electron-phonon interaction.

Before moving to the detailed analysis of the results, we emphasize that the RPA kind of scheme used here is generally valid in describing high electron density liquids. In doped semiconductors, an actual density of itinerant charge carriers is usually quite low in comparison to metals, albeit due to the smallness of effective masses and large values of effective Bohr radii, effective carrier densities may be even larger than in metals, making many semiconductors a suitable environment to implement the RPA \cite{mahan}. The use of the RPA for the phonon propagator is further supported by Migdal's theorem \cite{migdal1958}, which justifies the omission of electron-phonon vertex corrections for sufficiently high electron densities. Moreover, as long as Migdal's theorem may be applied, it may be argued that electron self-energy contributions bring only quantitative corrections to spectra, for example through the (weak) renormalization of effective masses or damping of excitations \cite{caruso2016}, while qualitatively no new features should be expected.

\section{Adiabaticity parameter and limiting behaviors\label{SecIII}}

A normal electron liquid involves the electron-hole pair excitations and the collective plasmon excitation. In the presence of the EPI, those excitations may strongly mix with the phonon degree of freedom. This complex mixing is captured by Eqs.~\eqref{eq:totaldielectricfunction} and \eqref{eq:RPApolarization}, accounting for the full dynamical treatment of the total dielectric function and the phonon self-energy in the presence of electrons coupled to the lattice. However, for a better understanding of the interplay between the lattice and the electron subsystem, it is particularly useful to consider limiting cases first. 

The plasmon frequency $\Omega^\infty_{PL}$ sets the frequency scale for the electron subsystem, similarly to the way the phonon frequency $\omega_{LO}$ characterizes the lattice subsystem. Thus, it is natural to introduce the adiabaticity parameter, measuring the ratio of those two frequency scales $\kappa = \omega_{LO}/\Omega^\infty_{PL}$. Depending on the value of $\kappa$, three different regimes may be distinguished, followed by three limiting choices for the electron dielectric function $\varepsilon_{RPA}(\mathbf{q}, \omega)$ in Eqs.~\eqref{eq:totaldielectricfunction} and \eqref{eq:RPApolarization}.

The large-$\kappa$ case corresponds to the antiadiabatic regime, when the plasmon frequency is significantly lower than that of the phonon. Apart from the polaronic effects \cite{Barisic,Bonca,Feinberg,Proville,mishchenko2002}, the phonon remains unrenormalized, since the slow electron plasma oscillations, $\Omega_{PL}^2\propto n$, cannot influence the fast lattice vibrations. This readily follows from Eq.~\eqref{eq:RPApolarization}, by noting that only the high-frequency part of the electron dielectric function $\varepsilon_{RPA}(\mathbf{q}, \omega\approx \infty)\to 1$ contributes in the frequency window situated around the phonon frequency.

By increasing the electron density the resonant regime sets in, when the electron density is tuned so that the plasmon frequency approximately matches that of the phonon, $\kappa \approx 1$. In such situations, the full frequency dependence of the electron dielectric function should be considered.

Lastly, for small $\kappa$, corresponding to the adiabatic limit, the density of itinerant charges is high and electron degrees of freedom are much faster than the phonon. Correspondingly, the electron dielectric function in Eqs.~(\ref{eq:totaldielectricfunction}) and (\ref{eq:RPApolarization}) may be approximated by its static value $\varepsilon_{RPA}(\mathbf{q}, \omega\approx 0)$. This suggests that the correct way of approximating the total dielectric function or the phonon self-energy is to first determine their dynamical properties through the electron dielectric function $\varepsilon_{RPA}(\mathbf{q}, \omega)$ depending on the adiabaticity parameter $\kappa$, and only after that eventual approximations on their wave vector dependence can be made.

\subsection{Resonant regime}

The resonant regime naturally serves as a starting playground for studying the phonon-plasmon mixing, since then the frequencies of two excitations are nearly degenerate.
Specially, in the long-wavelength limit, the physics of the resonant regime reduces solely to the phonon-plasmon coupling due to the absence of the continuum and the corresponding polaronic effects. In particular, the dynamical long-wavelength limit of electron dielectric function takes the form $\varepsilon_{RPA}(q\to 0 , \omega) = 1 -  (\Omega^\infty_{PL}) ^2/\omega^2$, yielding for the phonon self-energy

\begin{equation} \label{eq:RPApolarizationPlasmon}
	\Pi(q\to 0,\omega) =\frac{\hbar\Omega^\infty_{PL} |M_\mathbf{q}|^2}{2v^{\infty}_\mathbf{q}}\frac{2\hbar \Omega^\infty_{PL}}{\left( \hbar\omega\right)^2 -\left( \hbar\Omega^\infty_{PL}-i\eta\right)^2 }\;.
\end{equation}

\noindent The second factor on the right-hand side of  Eq.~\eqref{eq:RPApolarizationPlasmon} has the form of the free boson propagator, while the first factor may be interpreted as the effective matrix element of phonon-plasmon coupling. 

By inserting Eq.~(\ref{eq:RPApolarizationPlasmon}) into the Dyson equation in Eq.~(\ref{eq:dyson}) and by looking for the poles of the phonon propagator, the biquadratic equation is obtained, describing the coupling of two boson modes. Its solutions are given by

\begin{widetext}
	\begin{equation} \label{eq:plasmonphonon}
		\begin{split}
			2\omega_{\pm}^2 =\omega_{LO}^2+\left[ \Omega_{PL}^\infty\right] ^2
			\pm\sqrt{\left( \omega_{LO}^2-\left[ \Omega_{PL}^\infty\right] ^2\right)^2 + 16|\tilde{M}_\mathbf{q}|^2\omega_{LO} \Omega_{PL}^\infty /\hbar^2}\;,
		\end{split}	
	\end{equation}
\end{widetext}

\noindent corresponding to the frequencies of the collective excitations of the coupled phonon-plasmon system. The same solutions are obtained from the zeros of Eq.~\eqref{eq:totaldielectricfunction}, assuming in Eq.~\eqref{eq:plasmonphonon} the polar coupling given by Eq.~\eqref{eq:polarmatrixelement}. Here, it should be stressed that in describing the phonon-plasmon coupled system via the total dielectric function, Eq.~\eqref{eq:totaldielectricfunction}, the polar coupling is explicitly assumed. On the other hand, the approach involving the phonon propagator allows for the general type of EPI matrix elements $M_\mathbf{q}$.

Although Eq.~\eqref{eq:plasmonphonon} is strictly speaking obtained in the resonant regime, it provides the excitations' frequencies of the phonon-plasmon coupled system around $q \approx 0$, irrespectively of the adiabaticity parameter. In particular, in the antiadiabatic $\kappa \gg 1$ regime, two solutions of Eq.~\eqref{eq:plasmonphonon} are $\omega_+ = \omega_{LO}$ and $\omega_- = \Omega_{PL}^0=\Omega_{PL}/\sqrt{\varepsilon_0}$. That is, the phonon frequency remains unchanged, while the plasmon gets screened by both the interband excitations and the lattice vibrations. As shown in Appendix \ref{weightsEELS}, the corresponding plasmon spectral weight in an EELS spectrum then equals

\begin{equation} \label{eq:weightplasmonzero}
	s_{\Omega^{0}_{PL}}\left( q\to0\right) = \frac{\hbar \Omega^{0}_{PL}}{2\varepsilon_{0}}\;.
\end{equation}

When the electron density is high enough so that the adiabatic regime is reached, the electron subsystem completely screens long-range interactions between ions. This reduces the frequency of the LO phonon to that of the TO phonon. Accordingly, for $\kappa\ll 1$, two solutions of Eq.~(\ref{eq:plasmonphonon}) are obtained, $\omega_- = \omega_{TO}$ and $\omega_+ = \Omega^\infty_{PL}$. Since only the interband excitations are fast enough to screen the plasmon, $\omega_+$ is characterized by $\varepsilon_\infty$. 

In order to obtain the momentum dependence of the $\omega_-$ mode for $\kappa\ll1$, one may take the static limit of the electron dielectric function in Eq.~\eqref{eq:RPApolarization}. As shown in Appendix \ref{3Dadiabatic}, with $\varepsilon_{RPA}(q\to 0 , \omega\approx 0) = 1 + q^2/q^2_{TF}$, where $q_{TF}$ is the Thomas-Fermi wave vector, one obtains 

\begin{equation}
	\omega_- = \sqrt{\omega_{TO}^2 + \omega_{pl}^2\frac{q^2}{q_{TF}^2}}\;.\label{qdep}
\end{equation}

\noindent For metals in the jellium model $\varepsilon_0$ diverges \cite{Tupitsyn}, and Eq.~\eqref{qdep} may be used to obtain $\omega_-$ by setting $\omega_{TO}=0$. The ionic plasma oscillations get screened, acquiring an acoustic dispersion $\omega_-/\omega_{pl}=q/q_{TF}$, with $\omega_{pl}=\omega_{LO}$.

\section{Results}

The behaviors discussed in Sec. \ref{SecIII} correspond to the long-wavelength limit, with nothing said about the large-$q$ behaviors or the character of the corresponding excitations. Without the electron-hole continuum in Eq.~(\ref{eq:plasmonphonon}), the damping effects are absent as well. To overcome these limitations, by preserving the full momentum and frequency dependence in Eqs.~(\ref{eq:EELSspectrum}) and~(\ref{eq:RPApolarization}), we investigate in detail the structure of EELS spectra and phonon spectral functions, along with the distribution of spectral weights among different excitations. Our results show that in addition to the adiabaticity parameter, the EPI strength is essentially important  for the shape of spectra of phonon-plasmon coupled systems, motivating us to discuss regimes, with the experimentally relevant, the weak and the strong EPI separately. From the technical point of view, we use the analytical expression for the 3D Lindhard function $\chi_0(\mathbf{q},\omega)$ \cite{mihaila2011}, which simplifies the numerical work. 

\subsection{Weak coupling}

All our calculated spectra correspond to actual materials. As a first model of a bulk polar semiconductor, we consider the frequently studied GaAs. With the effective mass $m^*=0.0657 m$, the energy of the LO phonon $\hbar\omega_{LO} = 36.77$ meV, and the dielectric constants $\varepsilon_\infty = 10.9$ and $\varepsilon_0 = 12.83$ \cite{tempere2001}, it qualifies as a material with a weak polar coupling $\alpha\approx 0.07$.

\begin{figure*}
	\centering
	\includegraphics[width=\textwidth]{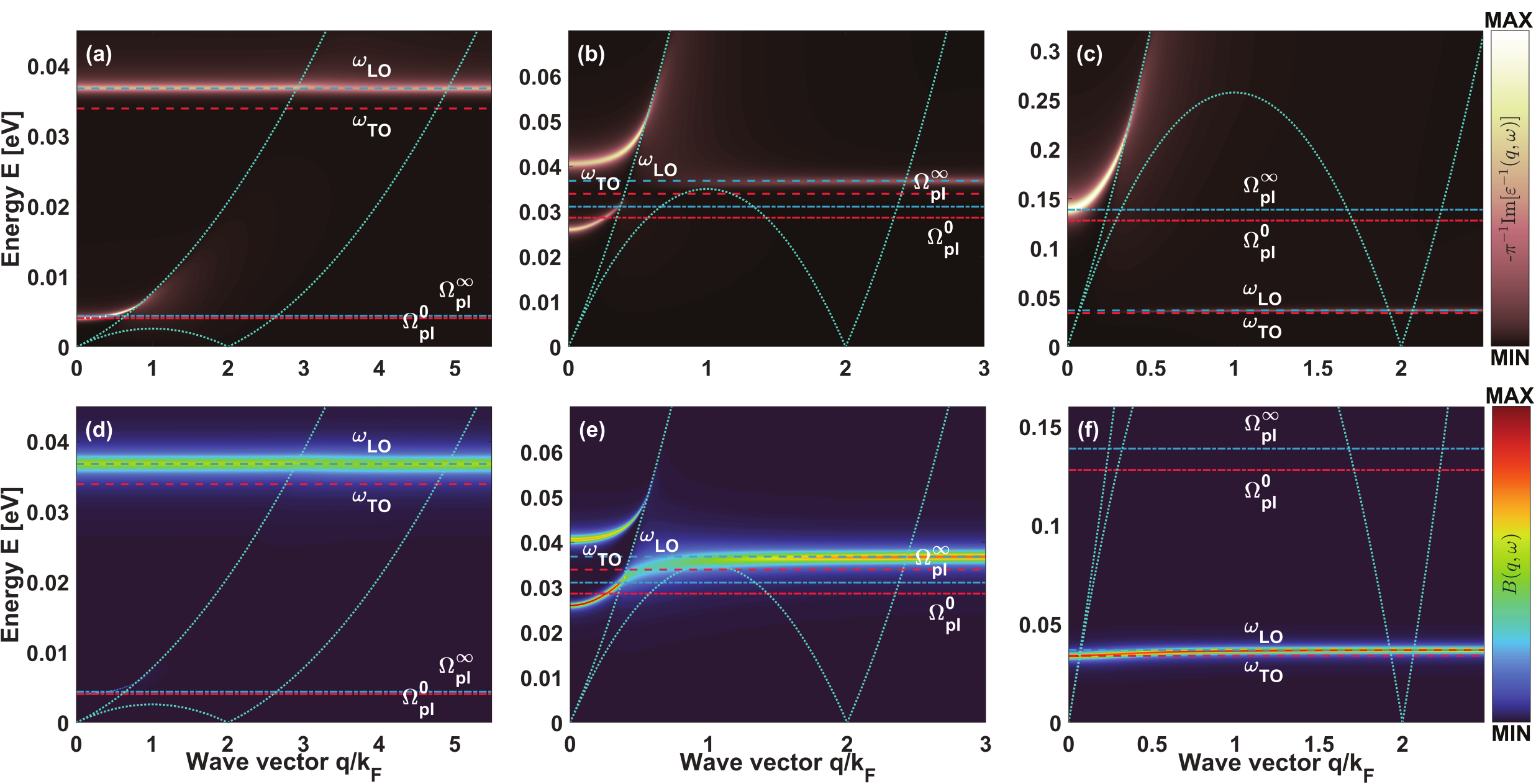}
	\renewcommand{\figurename}{Fig.}
	\caption{EELS spectra (upper row) and phonon spectral functions (corresponding to the LO phonon) (lower row) of the bulk GaAs for three different electron densities $n$. The first column corresponds to $n = 10^{16}$ cm$^{-3}$, the second to $n = 5\cdot 10^{17}$ cm$^{-3}$, and the third to $n = 10^{19}$ cm$^{-3}$, depicting the antiadiabatic $(\kappa \gg1)$, resonant $(\kappa \approx1)$, and adiabatic $(\kappa \ll1)$ regime, respectively. Note that MAX on the intensity scale takes a different absolute value for each of the panels.}
	\label{fig:3Dweak}
\end{figure*}

\begin{figure*}
	\centering
	\includegraphics[width=\textwidth]{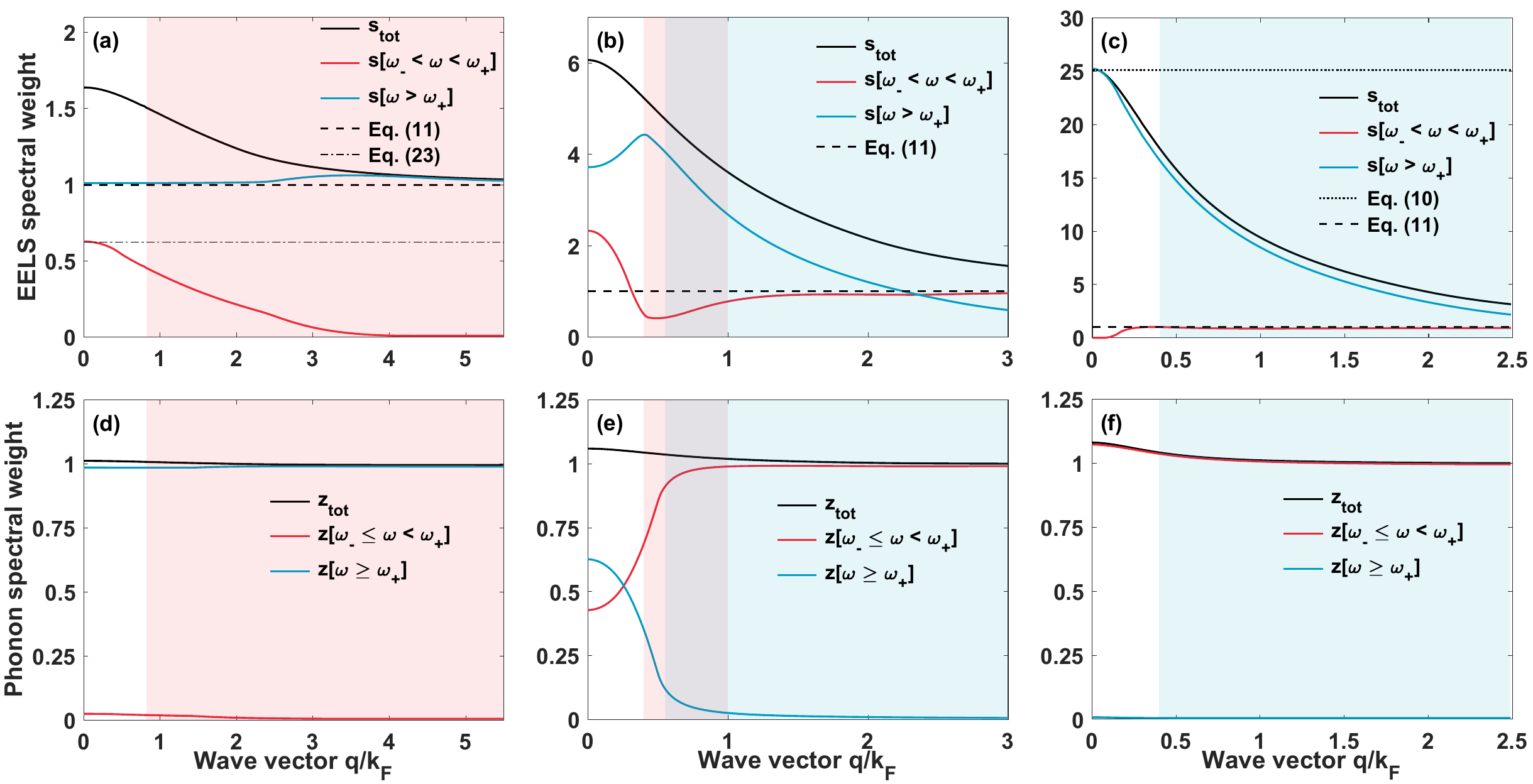}
	\renewcommand{\figurename}{Fig.}
	\caption{Integrated EELS spectra (upper row) and integrated phonon spectra (lower row) shown in Fig. \ref{fig:3Dweak}.}
	\label{fig:3Dweakintegrated}
\end{figure*}

\subsubsection{Spectral functions}

In Fig. \ref{fig:3Dweak}, we show the EELS spectra, Figs.~\ref{fig:3Dweak}(a)-\ref{fig:3Dweak}(c), and the phonon spectral functions of the LO phonon, Figs.~\ref{fig:3Dweak}(d)-\ref{fig:3Dweak}(f), of the doped bulk GaAs for three different electron densities $n =10^{16}$ cm$^{-3}$ $(k_F = 6.6\cdot 10^{-3}\;\angstrom^{-1})$, $5\cdot 10^{17}$ cm$^{-3}$ $(k_F = 2.4\cdot 10^{-2}\;\angstrom^{-1})$, and  $ 10^{19}$ cm$^{-3}$ $(k_F = 6.6\cdot 10^{-2}\;\angstrom^{-1})$. In all the figures, the red and the blue dashed lines denote the phonon frequencies $\omega_{TO}$ and $\omega_{LO}$, respectively. The plasmon frequencies $\Omega_{PL}^0$ and $\Omega_{PL}^\infty$ in the long-wavelength limit are denoted by the dot-dashed red and the dot-dashed blue lines, respectively, while the boundaries of the electron-hole continuum are denoted by the turquoise dotted lines.

From Figs.~\ref{fig:3Dweak}(a) and \ref{fig:3Dweak}(d), for $q=0$ we see two well-defined excitations at frequencies $\Omega^0_{PL}$ and $\omega_{LO}$, indicating clearly that the electron density $n = 10^{16}$ cm$^{-3}$ corresponds to the antiadiabatic regime. As $q$ increases, the LFE follows a plasmon-like dispersion and gets Landau damped upon entering the continuum. The HFE stays a well-defined excitation of constant frequency $\omega_{LO}$ up to the highest values of $q$ shown. 

Upon increasing the electron density, the resonant regime is reached, shown in Figs.~\ref{fig:3Dweak}(b) and~\ref{fig:3Dweak}(e). The strong level repulsion of coupled excitations is evident for smaller momenta $q<k_F$, because of which neither of two excitations in Figs.~\ref{fig:3Dweak}(b) and~\ref{fig:3Dweak}(e) exhibit long-wavelength limiting behaviors denoted by horizontal lines. With increasing $q$, both excitations develop a considerable dispersion and get strongly Landau damped by the continuum. For $q\gtrsim k_F$, a well-defined collective excitation emerges again, which frequency approaches the LO phonon frequency $\omega_{LO}$ and whose lifetime becomes longer as $q$ increases.

For the highest electron density, the EELS spectrum shows only the HFE for low $q$, Fig.~\ref{fig:3Dweak}(c). The LFE is, however, well captured by the phonon spectral function in Fig.~\ref{fig:3Dweak}(f), with exactly the frequency of the TO phonon for $q\approx0$, suggesting the adiabatic behavior of the system for the corresponding density. The absence of the LFE in the long-wavelength limit in the EELS spectrum is in accordance with the vanishing spectral weight at $\omega_{TO}$, when, as shown in Appendix \ref{weightsEELS}, the total dielectric function diverges. In Fig. \ref{fig:3Dweak}(c), the HFE evidently follows the plasmon dispersion $\Omega_{PL}^\infty$, unaffected by the phonon and gets Landau damped in the continuum. As seen from Fig.~\ref{fig:3Dweak}(f), the frequency of the LFE continuously increases from $\omega_{TO}$ to $\omega_{LO}$ and remains a well-defined excitation for all momenta, although weakly damped upon entering the continuum [the large energy scale set by the HFE partially hinders these details in Fig.~\ref{fig:3Dweak}(f)].

\subsubsection{Integrated spectra}

In order to get a better insight into the nature of excitations in Fig.~\ref{fig:3Dweak} as a function of $q$, in Fig.~\ref{fig:3Dweakintegrated} we consider the corresponding integrated EELS spectra (\ref{eq:eelsintegrated}) and the phonon spectral weight (\ref{eq:phononintegrated}) over the relevant frequency ranges. In particular, aside from the total spectral weight, we have considered spectral weights in the two specific frequency regions. The first corresponds to the frequency window $\omega_-<\omega<\omega_+$, while the second to $\omega > \omega_+$ \cite{comment_integration}. This should provide an estimation of the spectral weights $s_\pm$ and $z_\pm$, even in the presence of stronger damping or a limited experimental resolution. In Fig.~\ref{fig:3Dweakintegrated}, the spectral weights corresponding to the limiting behaviors, Eqs.~(\ref{eq:weightplasmoninfinity}), (\ref{eq:weightLO}), and \eqref{eq:weightplasmonzero}, are indicated by straight lines as well, normalized to the spectral weight at $\omega=\omega_{LO}$. The momenta for which in Fig.~\ref{fig:3Dweak} the LFE and the HFE get damped are shaded by the red and the blue color, respectively. These two shaded areas overlap in Figs.~\ref{fig:3Dweakintegrated}(b) and~\ref{fig:3Dweakintegrated}(e), i.e., in the resonant regime. 

\begin{figure*}
	\centering
	\includegraphics[width=\textwidth]{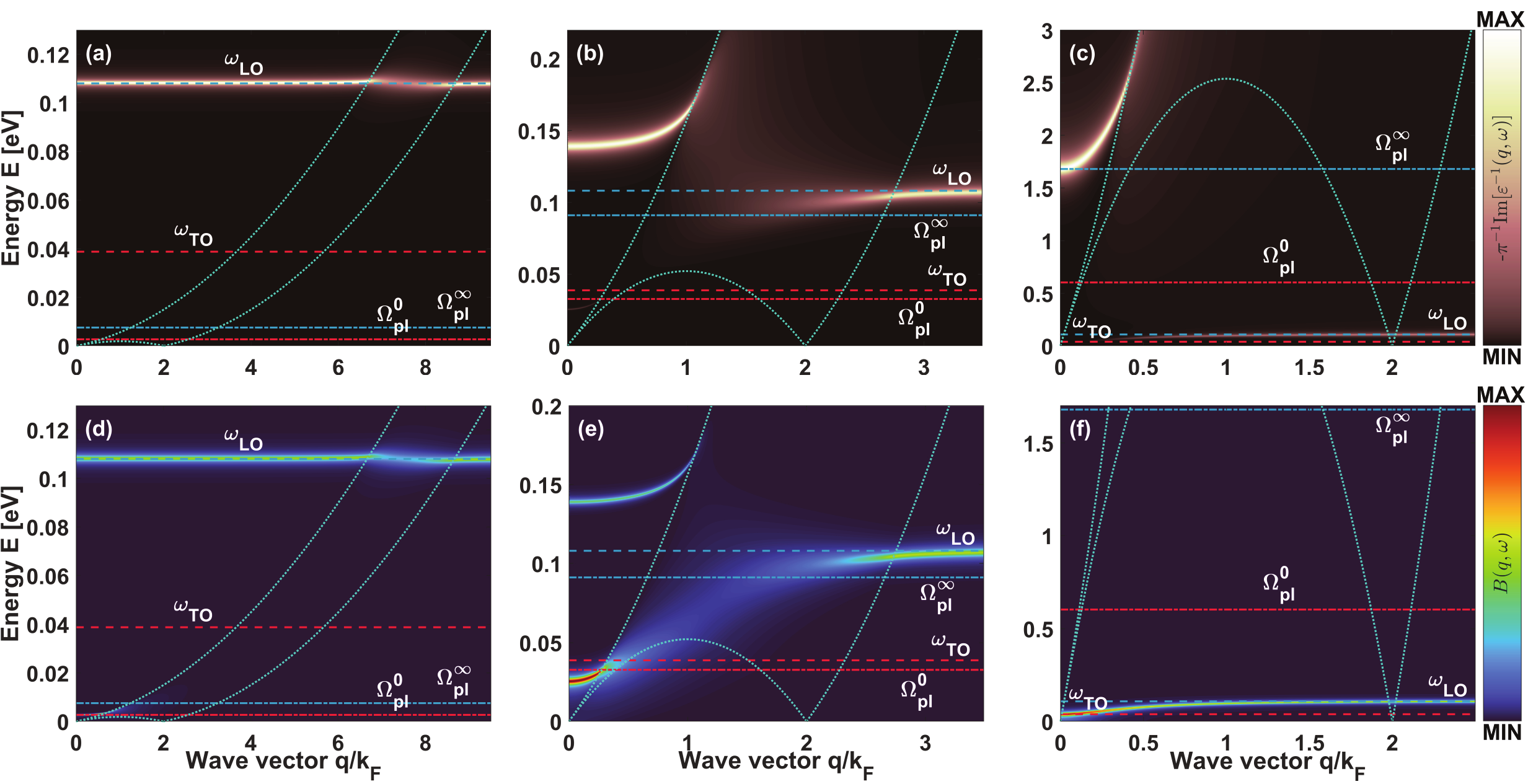}
	\renewcommand{\figurename}{Fig.}
	\caption{EELS spectra (upper row) and phonon spectral functions (corresponding to the LO $E_u$ phonon) (lower row) of the bulk anatase TiO$_2$ for three different electron densities $n$. The first column corresponds to $n =  10^{17}$ cm$^{-3}$, the second to $n = 1.47\cdot10^{19}$ cm$^{-3}$, and the third to $n = 5\cdot10^{21}$ cm$^{-3}$, depicting the antiadiabatic $(\kappa \gg1)$, resonant $(\kappa \approx1)$, and adiabatic $(\kappa \ll1)$ regime, respectively. Note that MAX on the intensity scale takes a different absolute value for each of the panels. To emphasize the role of electron-phonon coupling strength, the resonant regime is shown for the same value of $\kappa$ as in Fig.~\ref{fig:3Dweak}.}
	\label{fig:3Dstrong}
\end{figure*}

\begin{figure*}
	\centering
	\includegraphics[width=\textwidth]{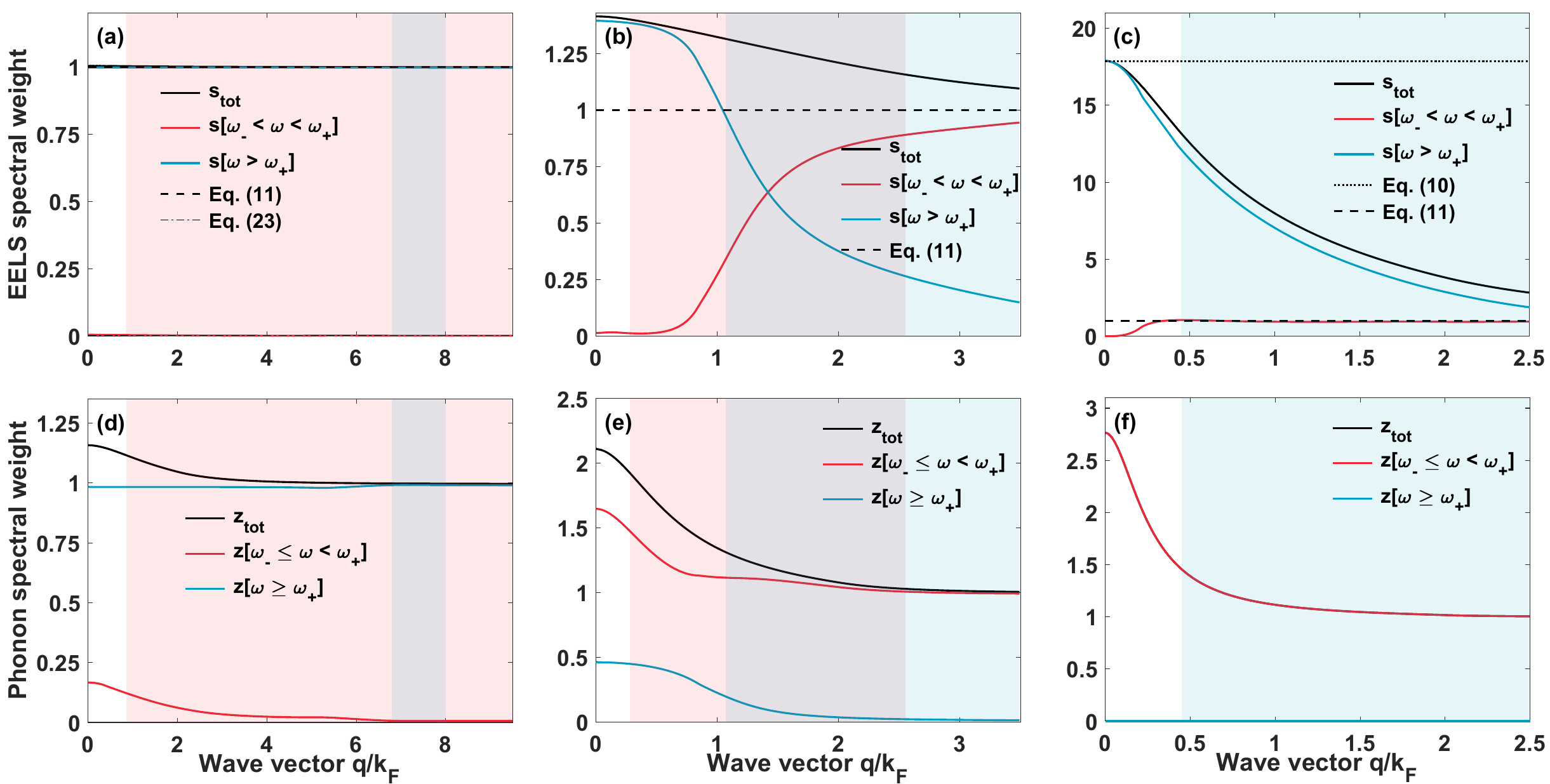}
	\renewcommand{\figurename}{Fig.}
	\caption{Integrated EELS spectra (upper row) and integrated phonon spectra (lower row) shown in Fig. \ref{fig:3Dstrong}.}
	\label{fig:3Dstrongintegrated}
\end{figure*}

For small momenta $q$, it is clear from Fig.~\ref{fig:3Dweakintegrated}(a) that, in the antiadiabatic regime, both the LFE and the HFE contribute significantly to the total EELS spectral weight, matching perfectly the predictions of Eqs.~\eqref{eq:weightplasmonzero} and~\eqref{eq:weightLO}, respectively. For larger $q$, the spectral weight of the LFE vanishes, while in the HFE case it stays roughly constant well described by Eq.~(\ref{eq:weightLO}), with a slight enhancement for $q\geq 3k_F$ that should be attributed to the electron-hole continuum $s_{e-h}$. Thus, the plasmon and the phonon character of the LFE and the HFE, respectively, is unquestionable for the antiadiabatic case. This is further confirmed by Fig.~\ref{fig:3Dweakintegrated}(d), with almost all the phonon spectral weight being associated with the HFE.

On the contrary, for the highest density case shown in Fig.~\ref{fig:3Dweakintegrated}(c), the HFE  spectral weight is almost purely plasmon-like. As predicted by  Eq.~(\ref{eq:weightplasmoninfinity}), it accounts for almost all of the total spectral weight at small momenta.  For higher momenta, the spectral weight in the frequency window $\omega > \omega_+$ should rather be attributed to the electron-hole continuum. On the other hand, the LFE spectral weight may be completely ascribed to the phonon subsystem. Indeed, the LFE contribution to the EELS spectrum appears in Fig.~\ref{fig:3Dweakintegrated}(c), when LFE approaches $\omega_{LO}$, due to the electron scattering by the LO vibrations, as described by Eq.~\eqref{eq:weightLO}. At the same time, the phonon spectral function in Fig.~\ref{fig:3Dweakintegrated}(f) is fully dominated by the LFE.

The results become slightly more difficult for interpretation in the resonant regime, Figs.~\ref{fig:3Dweakintegrated}(b) and~\ref{fig:3Dweakintegrated}(e). For small momenta, both collective excitations involve a strong mixture of the phonon and the plasmon component, signaling strongly hybridized modes. However, as seen from Fig.~\ref{fig:3Dweakintegrated}(e), for $q\geq0.4k_F$, the dominant character of excitations is unambiguous. In particular, the LFE is dominated by the phonon, while the HFE with the plasmon component.

The common property for all three phonon spectra in Fig.~\ref{fig:3Dweak} is a phonon production effect, manifested as a small increase of the total phonon spectral weight around $q\approx 0$. For larger $q$, this additional phonon spectral weight vanishes and the total spectral weight approaches the value given by Eq.~\eqref{eq:sumrule}. As we shall show, the phonon production contributions may become very large for the stronger EPI, with a very different origin depending on the adiabaticity parameter.

\subsection{Strong coupling}

In contrast to III-IV semiconductors, bulk transition metal oxides may host much stronger polar couplings. As a representative system with a significant EPI, we take anatase TiO$_2$, whose electrons upon doping the conduction band, characterized by the effective mass $m^* = 0.42\;m$ \cite{hitosugi2008}, couple to a LO $E_u$ phonon with the energy $\hbar\omega_{LO} = 108$ meV \cite{moser}. The corresponding dielectric functions read $\varepsilon_\infty = 5.82$ and $\varepsilon_0 = 45.1$ \cite{gonzalez1997}, resulting in the much larger EPI constant $\alpha \approx 1.09$ than in GaAs. From the experimental point of view, TiO$_2$ is very appealing, since $\omega_{LO}$ and Fermi wave vectors $k_F$ corresponding to relevant electron densities are a few times larger than for GaAs, making it more suitable for experiments with low energy and wave vector resolutions.

\subsubsection{Spectral functions}

In Fig. \ref{fig:3Dstrong}, we show the EELS spectra, Figs. \ref{fig:3Dstrong}(a)--\ref{fig:3Dstrong}(c), and the phonon spectral functions corresponding to the LO $E_u$ phonon, Figs. \ref{fig:3Dstrong}(d)--\ref{fig:3Dstrong}(f), of the bulk anatase TiO$_2$, considering three different electron densities $n =  10^{17}$ cm$^{-3}$ $(k_F = 1.4\cdot 10^{-2}\;\angstrom^{-1})$, $ 1.47\cdot 10^{19}$ cm$^{-3}$ $(k_F = 7.6\cdot 10^{-2}\; \angstrom^{-1})$, and $5\cdot10^{21}$ cm$^{-3}$ $(k_F = 0.53\; \angstrom^{-1})$. As in the already discussed weak-coupling case of GaAs, this choice of electron densities corresponds to the antiadiabatic, the resonant, and the adiabatic regime, respectively. The structure of spectra remains overall similar to that in the weak-coupling limit. However, it should be immediately emphasized that in the antiadiabatic and the resonant regime the excitations get much strongly damped within the continuum. Additionally, in the antiadiabatic regime, the HFE develops a visible kink as it enters the continuum, which may well be seen in Figs.~\ref{fig:3Dstrong}(a) and~\ref{fig:3Dstrong}(d).

\subsubsection{Integrated spectra}

Because of the strong damping, which completely blurs some parts of the EELS and the phonon spectra, for stronger couplings it is particularly useful to analyze the integrated spectra shown in Fig. \ref{fig:3Dstrongintegrated}. From Fig. \ref{fig:3Dstrongintegrated}(a), corresponding to the antiadiabatic regime, it is evident that the total spectral weight of the EELS spectrum is dictated by the constant spectral weight of HFE due to the strong EPI, Eq.~(\ref{eq:weightLO}), pointing to the phonon character of excitation, well supported by Fig. \ref{fig:3Dstrongintegrated}(d).

In the resonant regime, the spectral weight of HFE continues to dominate the EELS spectrum for small $q$, albeit higher spectral weight is confined within it than predicted by Eq.~(\ref{eq:weightLO}), suggesting  the appreciable plasmon component in addition to the phonon one. In the adiabatic regime, similarly to the weak-coupling case, the plasmon-like spectral weight of the HFE dominates the EELS spectrum for small $q$, while for larger momenta the total spectral weight is contributed by the LFE at $\omega_{LO}$, as described by Eq.~(\ref{eq:weightLO}), and the electron-hole continuum. The latter is also true in the resonant regime. In all the regimes, for small $q$, despite the change in the character of excitations, the HFE contributes much more to the total EELS spectral weight than the LFE.
Thus, unlike for weak couplings, the spectrum in the resonant regime in Figs.~\ref{fig:3Dstrong}(b) looks qualitatively similar to the spectra in the adiabatic regime in Figs.~\ref{fig:3Dweak}(c) and \ref{fig:3Dstrong}(c). Therefore, if the electron-phonon interaction strength is unknown, one may easily misinterpret to which regime does the EELS spectrum belong. However, if phonon measurements are available, the resonant regime may be identified without any ambiguity even for strong couplings.

Even bigger discrepancies between the weak and the strong coupling case are evident from the integrated phonon spectral functions, Figs. \ref{fig:3Dstrongintegrated}(d)-\ref{fig:3Dstrongintegrated}(f). The first striking result is that the additional phonon spectral weight, associated with the phonon production, is very large for small $q$. This large contribution in Fig. \ref{fig:3Dstrongintegrated} characterizes the LFE for all considered electron densities. However, the physical origin of this effect might be quite different depending on the adiabaticity parameter. In particular, in the antiadiabatic regime, the phonon production is apparently associated with the plasma oscillations. Indeed, by comparing the three curves in Fig. \ref{fig:3Dstrongintegrated}(d), we see that all the additional phonon spectral weight due to the phonon production clearly involves only the plasmon-like LFE, and not to the HFE, the latter having the predominantly phonon character. This additional phonon spectral weight scales sublinearly with the electron density, indicating that it involves a collective effect rather than being related to a simple polaronic dressing of individual electrons. On the other hand, in Fig.~\ref{fig:3Dstrongintegrated}(d), one as well observes a small contribution to the phonon production belonging to the electron-hole continuum. This stems from polaronic effects, corresponding to the phonon dressing of itinerant electrons, contributing to the Huang scattering \cite{Huang}.  

In the adiabatic limit, the electrons are faster and denser, for small $q$ almost fully screening the LO phonons. For this reason, $\omega_{TO}$ instead of $\omega_{LO}$ defines the small-$q$ LFE frequency in Fig.~\ref{fig:3Dstrong}(f). As $q$ increases, the LFE frequency approaches the LO phonon frequency, while the effects of the phonon production weaken. Our analysis of this additional phonon spectral weight in Fig.~\ref{fig:3Dstrongintegrated}(f), with details presented in Appendix~\ref{SecPPSS}, confirms that in the adiabatic limit the LFE should be interpreted as the pure harmonic excitation of the lattice subsystem; i.e., in Fig.~\ref{fig:3Dstrongintegrated}(f), as a function of $q$, the values obtained for the phonon production scale with the LFE frequency exactly as expected for the adiabatic phonon softening effect. For softer phonons with the LFE frequency the space uncertainty of lattice vibration increases, which through Eq.~\eqref{PHsumrule} explains the phonon production observed in Fig.~\ref{fig:3Dstrongintegrated}(f).

\section{Conclusions}

Our study systematically analyzes and compares features of EELS spectra and spectral functions corresponding to a LO phonon of 3D doped polar semiconductors.  The results are obtained in the zero-temperature limit when the spectra are the sharpest, by referencing to actual materials, namely GaAs and anatase TiO$_2$. While for the latter the EPI is strong, the former belongs to the weak-EPI limit. Thus, the comparison of these two cases permits us to identify the most important spectral features that depend on the strength of EPI, in parallel with the commonly studied influence of the electron density, i.e., the adiabaticity parameter.

In the adiabatic limit, the frequency of LFE smoothly changes from $\omega_ {TO}$ to $\omega_{LO}$, as the electron screening becomes ineffective for $q\gtrsim k_F$, while the HFE follows the plasmon dispersion $\Omega_{PL}^{\infty}$ before being completely damped by the continuum. The phonon-like LFE, on the other hand, is a well-defined excitation across the whole electron-phonon continuum, irrespective of the EPI strength. In the antiadiabatic limit, however, the phonon remains unscreened for all momenta, exhibiting the kink which gradually evolves into the continues LFE as the electron density increases. While the plasmon-like LFE gets completely Landau damped for all couplings, the broadening of the phonon-like HFE in the electron-hole continuum becomes significant only for strong EPIs. This dramatic influence of the EPI strength on damping is persistent in the resonant regime as well. Specifically, while the LFE for weak couplings behaves as a well-defined excitation through almost the whole continuum, for the strong EPIs it gets completely damped. The latter holds true for the HFE irrespective of the EPI strength, meaning that the existence of well-defined collective excitations over a broad range of momenta of the order of magnitude of the Fermi wave vector is heavily influenced by both the adiabaticity parameter and the EPI strength.

Apart from dictating the damping within the continuum, the EPI strength significantly influences the distribution of the EELS spectral weight among the excitations. In particular, for strong EPIs the HFE accounts for almost all the EELS spectral weight in the long-wavelength limit irrespective of the adiabaticity parameter, which appears as a robust feature of the strong-coupling regime. For the weak EPIs the same is true only in the adiabatic limit. In the antiadiabatic and the resonant regime, the EELS spectral weight is rather approximately equally redistributed among the LFE and the HFE, which opens the possibility of estimating the EPI strength from (integrated) EELS spectra, even from data with very limited energy resolutions. 

On the other hand, the study of phonon spectral weight emphasizes that the additional insight on the character of excitations may be obtained by identifying the phonon production contribution. In particular, in the adiabatic limit, the calculated phonon production confirms that the LFE is associated with the harmonic lattice vibrations, softened by the electron screening. On contrary, in the antiadiabatic limit, the phonon production effect is of a different origin, being associated with the LFE plasmon-like mode, indicating that due to the EPI the cloud of virtual phonons accompanies the plasma oscillations.  

As a final remark, our findings suggest that EELS measurement is in principal a powerful experimental tool to detect the HFE's dispersion, while the neutron inelastic scattering excels in capturing the LFE's dispersion. This highlights the complementarity of the EELS and the phonon spectra, and the advantages of studies of phonon-plasmon coupled systems when the both are experimentally accessible.

\begin{acknowledgments}
	Useful discussions with V. Despoja and D. Novko are acknowledged. J. K. acknowledges the support of the Croatian Science Foundation Project IP-2016-06-7258. O.S.B. acknowledges the support by the QuantiXLie Center of Excellence, a project co-financed by the Croatian Government and European Union through the European Regional Development Fund - the Competitiveness and Cohesion Operational Programme (Grant No. KK.01.1.1.01.0004).
\end{acknowledgments}

\begin{widetext}
\appendix

\section{EELS spectral weights\label{weightsEELS}}

In the long-wavelength limit, the total dielectric function of a heavily doped polar semiconductor in the RPA reads

\begin{equation}
\varepsilon(q\to 0,\omega) = \varepsilon_\infty - \frac{ \Omega_{PL}^2}{\omega^2} + \varepsilon_\infty\frac{\omega^2_{pl}}{\omega^2_{TO} -\omega^2}\;.
\end{equation}
Accordingly, one obtains

\begin{equation}
\begin{split}
\left[ \frac{\partial\varepsilon(q\to 0,\omega)}{\partial\omega}\right]^{-1}  = \frac{\hbar}{2}\frac{\left[ \omega^2_{TO} -\omega^2\right]^2\omega^3}{\Omega_{PL}^2\left[ \omega^2_{TO} -\omega^2\right]^2 + \varepsilon_\infty\omega^4\omega^2_{pl}}\;.
\end{split}
\end{equation}
It is now trivial to see that the EELS spectral weight of the TO phonon vanishes, $s_{\omega_{TO}}\left( q\to0\right) = \left.\left[ \frac{\partial\varepsilon(q\to 0,\omega)}{\partial\omega}\right]^{-1}\right| _{\omega=\omega_{TO}} = 0$.

Next, we evaluate the EELS spectral weights at plasmon frequencies $\Omega_{PL}^\infty$ and $\Omega_{PL}^0$, screened by dielectric constants $\varepsilon_{\infty}$ and $\varepsilon_{0}$, respectively, and the frequency of LO phonon $\omega_{LO}$,

\begin{equation}
\begin{split}
s_{\Omega^{\infty}_{PL}}\left( q\to0\right)= \left.\left[ \frac{\partial\varepsilon(q\to 0,\omega)}{\partial\omega}\right]^{-1}\right|_{\omega=\Omega_{PL}^\infty}  \stackrel{\Omega_{PL}\gg \omega_{pl}}{\approx}\frac{\hbar}{2}\frac{\left[ \Omega_{PL}^\infty\right]^3}{\Omega_{PL}^2}= \frac{\hbar\Omega_{PL}^\infty}{2\varepsilon_\infty} \;.
\end{split}
\end{equation}
Here, we assumed $\Omega_{PL}\gg \omega_{pl}$, which is satisfied in the adiabatic regime, where $\Omega_{PL}$ and $\omega_{pl}$ denote frequencies of the electronic and the ionic plasma, respectively. On the other hand, in the antiadiabatic regime both $\Omega_{PL}^0\ll \omega_{TO}$ and $\Omega_{PL}\ll \omega_{pl}$ hold, resulting in

\begin{equation}
\begin{split}
s_{\Omega^{0}_{PL}}\left( q\to0\right) = \left.\left[ \frac{\partial\varepsilon(q\to 0,\omega)}{\partial\omega}\right]^{-1}\right|_{\omega=\Omega_{PL}^0}  &
\stackrel{\Omega_{PL}^0\ll \omega_{TO}}{\approx}\frac{\hbar}{2}\frac{\left[ \Omega_{PL}^0\right]^3}{\Omega_{PL}^2}=  \frac{\hbar\Omega_{PL}^0}{2\varepsilon_0}\;,
\end{split}
\end{equation}
and

\begin{equation}
\begin{split}
s_{\omega_{LO}}\left( q\to0\right)=\left.\left[ \frac{\partial\varepsilon(q\to 0,\omega)}{\partial\omega}\right]^{-1}\right|_{\omega=\omega_{LO}}  \stackrel{\Omega_{PL}\ll \omega_{pl}}{\approx}\frac{\hbar}{2}\frac{\omega^2_{LO}-\omega^2_{TO}}{\varepsilon_\infty \omega_{LO}}=\frac{\hbar}{2}\frac{\omega^2_{LO}-\omega^2_{LO}\frac{\varepsilon_\infty}{\varepsilon_0}}{\varepsilon_\infty \omega_{LO}}= \frac{\hbar\omega_{LO}}{2}\left( \frac{1}{\varepsilon_\infty}-\frac{1}{\varepsilon_0}\right) \;.
\end{split}
\end{equation}

\section{Long-wavelength dispersion of the phonon-like mode in the adiabatic regime\label{3Dadiabatic}}
To get the small $q$ dependence of phonon mode dispersion in the adiabatic regime, the static electron dielectric function $\varepsilon_{RPA}(q \to 0, \omega\approx 0)=1 + \frac{q_{TF}^2}{q^2}\Rightarrow 1 / \varepsilon_{RPA}(q \to 0, \omega\approx 0) \approx \frac{q^2}{q_{TF}^2}$ has to be taken in Eq.~\eqref{eq:RPApolarization} for the phonon polarization, with $q_{TF}$ the Thomas-Fermi wave vector. Now, by solving the Dyson equation for the LFE, corresponding to the phonon-like mode, with the polar coupling $M_\mathbf{q} = -i\sqrt{v_\mathbf{q}^{\infty}} \sqrt{\frac{\hbar \omega_{LO}}{2}} \sqrt{1 - \frac{\varepsilon_\infty}{\varepsilon_0}}$, one gets

\begin{equation}
\begin{split}
\omega_-^2&=\omega_{LO}^2 + 2\omega_{LO}\frac{|M_\mathbf{q}|^2}{v_\mathbf{q}^{(\infty)}}\left[ \frac{q^2}{q_{TF}^2}-1\right] = \omega_{LO}^2 \left\lbrace 1+\left[ 1-\frac{\varepsilon_\infty}{\varepsilon_0}\right]\left[ \frac{q^2}{q_{TF}^2}-1\right]\right\rbrace  \stackrel{LST}{=} \omega_{TO}^2+\frac{\omega_{pl}^2}{q_{TF}^2}q^2\;.
\end{split}	
\end{equation}

\section{Phonon production for squeezed states\label{SecPPSS}}

The Lehmann representation of the LO phonon Green's function is given by

\begin{equation}
D(\mathbf{q},\omega)=\sum_n|\langle0|(a_\mathbf{q}+a_\mathbf{-q}^\dagger)|n\rangle|^2\left(\frac{1}{\omega-(E_n-E_0)+i\eta}-
\frac{1}{\omega-(E_0-E_n)-i\eta}\right)\;,\label{DexactApp}
\end{equation}

\noindent where  $|0\rangle$ and $|n\rangle$ are the exact ground and the excited states of the (interacting) system, respectively. From Eq.~(\ref{DexactApp}) it is easy to check that the integrated LO phonon spectral weight satisfies Eq.~(\ref{PHsumrule}),

\begin{equation}
\int_{-\infty}^\infty -\frac{1}{2\pi}\IM D(\mathbf{q},\omega)d\omega=\sum_n|\langle0|(a_\mathbf{q}+a_\mathbf{-q}^\dagger)|n\rangle|^2
=\langle0|(1+2a_\mathbf{q}^\dagger a_\mathbf{q}+a_\mathbf{q} a_\mathbf{-q}+a_\mathbf{q}^\dagger a_\mathbf{-q}^\dagger)|0\rangle\;.\label{SumApp}
\end{equation}

Assuming that the ground state of the lattice subsystem is given by squeezed states of harmonic oscillators

\begin{equation}
|0\rangle=\Pi_\mathbf{q}\exp\left(\frac{1}{2}\left( \gamma_\mathbf{q}^*a_\mathbf{q}^2-\gamma_\mathbf{q} a_\mathbf{q}^{\dagger2}\right) \right)|\tilde 0\rangle\;,
\end{equation}

\noindent with $\gamma=|\gamma|e^{i\Theta}$ and $|\tilde 0\rangle$ the LO phonon vacuum, from Eq.~(\ref{DexactApp}) one obtains

\begin{equation}
\langle0|\hat x_\mathbf{q}\hat x_\mathbf{-q}|0\rangle
=\frac{\hbar}{2M\omega_{LO}}\left(\cosh^2(|\gamma_\mathbf{q}|)+\sinh^2(|\gamma_\mathbf{q}|)+2\cos\Theta\cosh(|\gamma_\mathbf{q}|)\sinh(|\gamma_\mathbf{q}|)\right)\;.
\label{PHsumruleApp}
\end{equation}

For $\Theta=\pi$, the  squeezed state is elongated along the real-space coordinate $x_\mathbf{q}$. Thus, assuming that the LO phonon is fully screened and that the zero-point motion is characterized by the TO frequency, one gets

\begin{equation}
\langle0|\hat x_\mathbf{q}\hat x_\mathbf{-q}|0\rangle=\frac{\hbar}{2M\omega_{LO}}e^{2|\gamma|}=\frac{\hbar}{2M\omega_{TO}}\;.
\end{equation}

\noindent For GaAs and TiO$_2$, $e^{2|\gamma|}=\omega_{LO}/\omega_{TO}=\sqrt{\varepsilon_0/\varepsilon_\infty}$ approximately equals $e^{2|\gamma|}\approx 1.08$ and $e^{2|\gamma|}\approx 2.78$, respectively.  Those are almost the same values of the total phonon spectral weights obtained in Figs.~\ref{fig:3Dweakintegrated}(f) and \ref{fig:3Dstrongintegrated}(f) for the soft $q\approx0$ phonon, whose frequency is very close to $\omega_{TO}$.

\end{widetext}


\begin{thebibliography}{99}
\bibitem{comment_integration} For example, to capture the total spectral weight in the HFE, $s_+$ or $z_+$, the integration cutoff is defined below $\omega_+$ such that it correctly reproduces the corresponding spectral weight in the long-wavelength limit. In particular, in the resonant regime a weakly momentum dependent integration cutoff has been taken.
\end{thebibliography}
\end{document}